\documentclass[%
 reprint,
superscriptaddress,
nofootinbib,
 amsmath,amssymb,
 aps,
prl,
]{revtex4-2}
\usepackage[dvipsnames]{xcolor}
\usepackage{graphicx}
\usepackage{dcolumn}
\usepackage{bm}
\usepackage{hyperref}

\usepackage{braket}
\usepackage[utf8]{inputenc} 
\usepackage{mathtools}
\usepackage[bottom]{footmisc}
\usepackage{hyperref}
\usepackage{amsfonts}
\usepackage{amsmath}
\usepackage{slashed}
\usepackage{amsthm}
\usepackage[normalem]{ulem}
\usepackage{soul}

\usepackage[english]{babel}

\begin{document}

\preprint{APS/123-QED}

\title{Topological Nature of Orbital Chern Insulators}

\author{Yueh-Ting Yao}
\affiliation{Department of Physics, National Cheng Kung University, Tainan, 70101, Taiwan}

\author{Chia-Hung Chu}
\affiliation{Department of Physics, National Cheng Kung University, Tainan, 70101, Taiwan}

\author{Arun Bansil}
\affiliation{Department of Physics, Northeastern University, Boston, Massachusetts, 02115, USA}
\affiliation{Quantum Materials and Sensing Institute, Northeastern University, Burlington, MA 01803, USA}

\author{Hsin Lin}
\email{nilnish@gmail.com}
\affiliation{Institute of Physics, Academia Sinica, Taipei, 115201, Taiwan}

\author{Tay-Rong Chang}
\email{u32trc00@phys.ncku.edu.tw}
\affiliation{Department of Physics, National Cheng Kung University, Tainan, 70101, Taiwan}
\affiliation{Center for Quantum Frontiers of Research and Technology (QFort), Tainan, 70101, Taiwan}
\affiliation{Physics Division, National Center for Theoretical Sciences, Taipei, 10617, Taiwan}

\date{\today} 

\begin{abstract}
Ground state topologies in quantum materials have unveiled many unique topological phases with novel Hall responses. Recently, the orbital Hall effect in insulators has suggested the existence of orbital Chern insulators (OCIs) in which the orbital angular momentum drives the Hall response. Studies on OCIs, however, have so far been restricted to valley-locked or spinful systems, but candidate materials for systematic studies of OCIs are lacking. Here we discuss a framework for investigating OCIs using the feature-spectrum topology approach. To characterize the ground-state topology in the orbital degree of freedom, we introduce the orbital Chern number and orbital-feature Berry curvature and demonstrate the bulk-boundary correspondence and orbital Hall response. We also uncover a parameter-driven topological phase transition, which would offer tunability of the OCIs. In this way, we identify monolayer blue-phosphorene (traditionally considered topologically trivial) as the primal ‘hydrogen atom’ of OCIs as a spinless, valley-free OCI material. Our study gives insight into the nature of orbital-driven topological phases and reveals a new facet of blue-phosphorene, and provides a new pathway for advancements in orbitronics and the discovery of novel topological materials.
\end{abstract}



\maketitle



\par Exploration of ground-state topology has led to the discovery of topological phases of quantum matter across various degrees of freedom\cite{hasan_colloquium_2010,bansil_colloquium_2016, qi_topological_2011}. These topological phases exhibit unique behaviors, such as the topologically protected boundary states and Hall conductivity plateaus within the insulating band gaps, spurring much theoretical and experimental research \cite{chang_experimental_2013,deng_quantum_2020,konig_quantum_2007,knez_evidence_2011,wu_observation_2018}. In two dimensions, the concept of Chern insulators (CIs) led to the discovery of the quantum anomalous Hall effect with chiral edge states\cite{haldane_model_1988, chang_experimental_2013, deng_quantum_2020}, while spin Chern insulators (SCIs) gave rise to the quantum spin Hall effect with helical edge states\cite{sinova_spin_2015,kane_z_2005,kane_quantum_2005,sheng_quantum_2006, bernevig_quantum_2006, fu_topological_2007,sinova_universal_2004,konig_quantum_2007,knez_evidence_2011,wu_observation_2018}. Chern-like invariants play a pivotal role in characterizing these topologically insulating phases\cite{thouless_quantized_1982,haldane_model_1988,kane_z_2005,kane_quantum_2005,sheng_quantum_2006, prodan_robustness_2009}, providing a fundamental framework for understanding effects of topology across charge and spin degrees of freedom and paving the way for the exploration of novel phases of quantum matter as well as advancements in electronics and spintronics applications. 

\par Beyond the charge and spin degrees of freedom, recent studies in orbitronics have unveiled the transverse response of current-carrying orbital angular momentum (OAM), referred to as the orbital Hall effect (OHE)\cite{atencia_orbital_2024, jo_spintronics_2024, choi_observation_2023, lyalin_magneto-optical_2023, lee_orbital_2021, gupta_harnessing_2025, bernevig_orbitronics_2005, go_orbitronics_2021, jo_gigantic_2018, go_intrinsic_2018, pezo_orbital_2023, go_orbital_2020, canonico_orbital_2020, bhowal_intrinsic_2020, cysne_disentangling_2021, cysne_orbital_2022, busch_orbital_2023, pezo_orbital_2022, ji_reversal_2024, costa_connecting_2023, yang_monopole-like_2023}. Experimentally, OHE has been observed in transition metals with weak spin-orbit coupling (SOC), where significant OHE signals have been detected through the magneto-optical response of orbital torque\cite{manchon_current-induced_2019, choi_observation_2023, lee_orbital_2021, lyalin_magneto-optical_2023, gupta_harnessing_2025}, highlighting the potential of OHE in enhancing power efficiency and nonvolatility in next-generation memories.\cite{gupta_harnessing_2025}. Theoretical studies of semiconductors have revealed orbital Hall conductivity (OHC) plateaus within their insulating band gaps\cite{canonico_two-dimensional_2020,canonico_two-dimensional_2020,canonico_orbital_2020, bhowal_intrinsic_2020, cysne_disentangling_2021, cysne_orbital_2022, busch_orbital_2023, pezo_orbital_2022, ji_reversal_2024, costa_connecting_2023, wang_topological_2024}. The presence of plateaus in OHC suggests the existence of Chern-like insulating phases in the orbital degree of freedom, leading to the concept of orbital Chern insulators (OCIs), also known as orbital Hall insulators\cite{canonico_two-dimensional_2020,canonico_orbital_2020}. The topological nature of OCIs can be independently characterized via the orbital Chern number. However, previous studies on OCIs have been limited to valley-localized states\cite{canonico_two-dimensional_2020,cysne_disentangling_2021} or descriptions involving entanglement with spin degrees of freedom via spin-orbit coupling\cite{wang_topological_2024}, see Table.~\ref{tab1}. Involvement of other degrees of freedom, obscures the topological nature of the OCIs, and a more fundamental framework for characterizing the OCIs and identifying viable materials candidates is important for understanding the nature of topology in the orbital degree of freedom.

\par In this work, we present a comprehensive framework for investigating OCIs as a Chern-like topological ground state in the orbital degree of freedom using the feature-spectrum topology approach\cite{wang_feature_2023, lin_spin-resolved_2024, yao_feature-energy_2024, wang_topological_2024}. The two-dimensional (2D) honeycomb pnictogen (group-5A elements) monolayers are shown to be promising candidate materials for realizing the OCIs, supported by density functional theory (DFT) calculations. Monolayer blue-phosphorene, for example, is found to be a valley-free, spinless OCI, despite its classification as topologically trivial in the existing literature\cite{ tani_topological_2022,ezawa_topological_2014}. To characterize the ground-state topology in the orbital degree of freedom, we introduce the orbital Chern number along with the orbital-feature Berry curvature. Importantly, we demonstrate that the bulk-boundary correspondence in OCIs manifests itself in the form of floating edge states, which are residual features of nontrivial states in the edge $\hat{L}_z$-feature spectrum, in addition to the presence of an OHC plateau within the bulk band gaps. We also uncover a parameter-driven topological phase transition, which would offer tunability of the OCIs. Our study provides new insights into the role of the orbital degree of freedom in topological phases of quantum matter and identifies monolayer blue-phosphorene as the primal ‘hydrogen atom’ of OCIs, opening new avenues for exploration of orbital-driven topological phenomena.

\begin{table}[t]
    \centering
    \begin{tabular}{c|c|c}
        \hline \hline
        \textbf{Materials} & Spinless & Valley-free \\
        \hline
        2\textit{H}-TMDs \cite{canonico_orbital_2020,bhowal_intrinsic_2020,cysne_orbital_2022,cysne_disentangling_2021,costa_connecting_2023} & $\times$ & $\times$ \\
        1\textit{T}-TMDs \cite{costa_connecting_2023} & $\times$ & $\vee$ \\
        Group-4A \cite{wang_topological_2024} & $\times$ & $\times$ \\
        Group-5A (This work) & $\vee$ & $\vee$ \\
        \hline \hline
    \end{tabular}
    \caption{Summary of material candidates for OCIs and whether they are restricted by spin and valley degrees of freedom.}
    \label{tab1}
\end{table}


\par Monolayer blue-phosphorene crystallizes in a buckled honeycomb structure within the space group $P\bar{3}m1$ (No.164)\cite{zhu_semiconducting_2014,gu_growth_2017,ezawa_minimal_2018} (Fig.~\ref{fig:01}(a)). DFT based band structure (without SOC) is shown in Figure.~\ref{fig:01}(b), see Supplementary Materials (SM) for details. An indirect band gap of approximately 2 eV is seen with the Chern number as well as the spin Chern number being zero\cite{ezawa_topological_2014, tani_topological_2022}.

\par Band inversion, which is a fundamental mechanism for inducing nontrivial topology, involves the exchange of wave function character between the valence and conduction states near the Fermi level\cite{bansil_colloquium_2016, zhang_topological_2009}. Although monolayer blue-phosphorene has been classified as a normal (trivial) insulator in the literature\cite{ tani_topological_2022,ezawa_topological_2014}, our analysis reveals the presence of band inversion between the magnetic quantum numbers ($\bar{m}$) of $l=1$ ($p$) orbitals. Specifically, the inversion is between the $\bar{m} = \pm1$ orbitals ($\frac{1}{\sqrt{2}}(p_{x}\pm i p_y)$) and $\bar{m} = 0$ ($p_z$) orbitals. The orbital decomposition in Fig.~\ref{fig:01}(b) shows hybridization between $p_z$ and $p_{x/y}$ orbitals, along with the band inversion at both $\Gamma$ and $M$ points. In contrast, Fig.~\ref{fig:01}(c) shows that in monolayer blue-phosphorene \textit{without buckling}, the $p_z$ orbitals are well-separated from the $p_{x/y}$ orbitals. The out-of-plane buckling is thus responsible for opening the inverted band gap between the in-plane $p_{x/y}$ and out-of-plane $p_z$ orbitals, see SM for details.  



\begin{figure}[t]
\centering
\includegraphics[width=\linewidth]{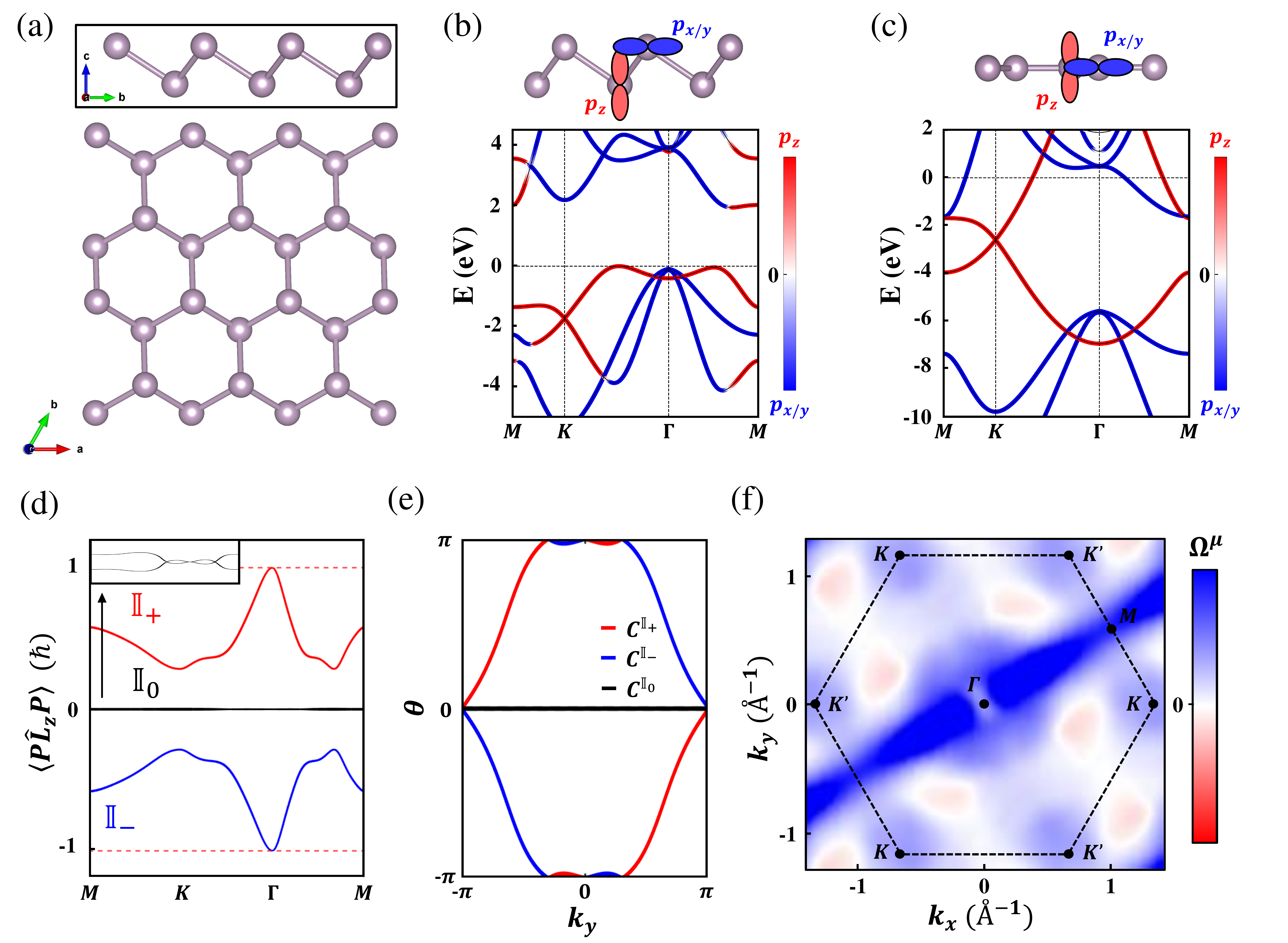}
\caption{(a) Crystal structure of honeycomb pnictogen monolayers, with the top and bottom panels showing the side and top views, respectively. (b-c) Band structure and orbital contributions in (b) buckled and (c) planar monolayer blue-phosphorene. (d) $\hat{L}_z$-projected feature spectrum of buckled monolayer blue-phosphorene, where the inset highlights the $\mathbb{I}_0$ sector. (e) $\hat{L}_z$-resolved Wilson loop for sectors. (f) $\hat{L}_z$-feature Berry curvature $\Omega^\mu$ of $\mathbb{I}_{\pm}$.
}
\label{fig:01}
\end{figure}

\par To investigate the nontrivial topology, we employ the feature-spectrum topology approach\cite{wang_feature_2023, lin_spin-resolved_2024, yao_feature-energy_2024, wang_topological_2024} using the OAM operator $\hat{L}_z = s_0\otimes\sigma_0\otimes\hat{l}_z$. Here, $s_0$ and $\sigma_0$ are $2\times2$ identity matrices acting on the spin and sublattice degrees of freedom, respectively, while $\hat{l}_z$ represents the z-component of the OAM operator for $p$-orbitals in the $|p_z\rangle,|p_x\rangle,|p_y\rangle$ basis\cite{canonico_orbital_2020, jo_gigantic_2018}:

\begin{equation}\label{eq:01}
\begin{split}
    \hat{l}_z = i\hbar \begin{pmatrix}
    0 & 0 & 0 \\
    0 & 0 & -1 \\
    0 & 1 & 0
    \end{pmatrix}.
\end{split}
\end{equation}

\noindent By projecting the valence electrons [Fig.~\ref{fig:01}(b)] using projection operator onto occupied states $P$, the feature operator is given by $P\hat{L}_zP$. The $\hat{L}_z$-projected feature spectrum, $\langle P\hat{L}_zP\rangle$, shown in [Fig.~\ref{fig:01}(d)], consists of three distinct sectors: $|l,\bar{m}\rangle=|1,0\rangle,|1,1\rangle,|1,-1\rangle$.
The $|1,0\rangle$ sector ($\mathbb{I}_0$) corresponds to the $p_z$-orbital, while the $|1,\pm 1\rangle$ sectors ($\mathbb{I}_\pm$) represent the $p_{x/y}$ orbitals. Assuming spherical symmetry, the relationship $\hat{L}_z|l,\bar{m}\rangle=\bar{m}\hbar|l,\bar{m}\rangle$ holds, where $\bar{m}\in\big\{0,\pm1\big\}$. At the $\Gamma$ point, the three sectors ($\mathbb{I}_{0,\pm1}$) in Fig.~\ref{fig:01}(d) reflect the spherical symmetry, which enforces the eigenvalues of $P\hat{L}_zP$ located at the $\hat{L}_z$ conserved values: $\bar{m}=0$ or $\pm1$. Away from the $\Gamma$ point, the spherical symmetry is broken by the lattice geometry, leading to a dispersive $\langle P\hat{L}_zP\rangle$ spectrum.



\par Topological invariants provide a powerful framework for identifying and classifying nontrivial topology in condensed matter systems across various degrees of freedom\cite{haldane_model_1988, thouless_quantized_1982, haldane_model_1988,kane_z_2005,kane_quantum_2005,sheng_quantum_2006, prodan_robustness_2009}. To identify the topological nature of monolayer blue-phosphorene, we introduce the feature-resolved orbital Chern number $C^\mu$\cite{wang_topological_2024} and define the $\hat{L}_z$-feature Berry curvature $\Omega^\mu(\mathbf{k})$ within the $P\hat{L}_zP$ feature space for each sector, where $\mu\in\big\{\mathbb{I}_0, \mathbb{I}_{+1}, \mathbb{I}_{-1}\big\}$ is the sector label. Figure~\ref{fig:01}(e) presents the feature-resolved orbital Chern numbers for three sectors, obtained via the $\hat{L}_z$-resolved Wilson loop. The Wilson loop windings indicate that $C^{\mathbb{I}_0}=0$, while $C^{\mathbb{I}_+}=1$ and $C^{\mathbb{I}_-}=-1$. Following a similar definition with spin Chern number\cite{prodan_robustness_2009}, the total orbital Chern number $C_L$ is given by $C_L=(C^{\mathbb{I}_+}-C^{\mathbb{I}_-})/2$\cite{cysne_disentangling_2021}. The resulting $C_L=1$ confirms that monolayer blue-phosphorene is an OCI. 

\par To further explore the geometric properties of OCIs, we introduce the $\hat{L}_z$-feature Berry curvature across the Brillouin zone. If the sectors are well-separated in the feature spectrum, the projection operator onto the occupied Bloch states can be decomposed as $P(\mathbf{k})=\bigoplus_{\mu}P^\mu(\mathbf{k})$, where $P^\mu(\mathbf{k})=\sum_{m\in\mu}^{N_{occ}^\mu}|\widetilde{u}^\mu_{mk}\rangle\langle\widetilde{u}^\mu_{mk}|$, with $|\widetilde{u}^\mu_{mk}\rangle$ are the feature Bloch states\cite{lin_spin-resolved_2024}(see SM for details) and $N_{occ}^\mu$ denotes the number of occupied states in sector $\mu$. This formulation allows geometric analysis of each sector through the $\hat{L}_z$-feature Berry curvature, which is derived from the geometrical form of the Berry curvature\cite{prodan_robustness_2009,mera_kahler_2021},
\begin{equation}\label{eq:02}
\begin{split}
\Omega^\mu(\mathbf{k})=tr(P^\mu(\mathbf{k})\left[\partial_{k_x}P^\mu(\mathbf{k}),\partial_{k_y}P^\mu(\mathbf{k})\right]).
\end{split}
\end{equation}
\noindent Figure~\ref{fig:01}(f) show the $\hat{L}_z$-feature Berry curvature for monolayer blue-phosphorene in the $\mathbb{I}_+$ and $\mathbb{I}_-$ sectors. The $\hat{L}_z$-feature Berry curvature is predominantly concentrated around the $\Gamma$ point and along the $\Gamma-M$ high-symmetry line, consistent with the band inversion behavior in Fig.~\ref{fig:01}(b). 



\par
\begin{figure}[t]
\centering
\includegraphics[width=\linewidth]{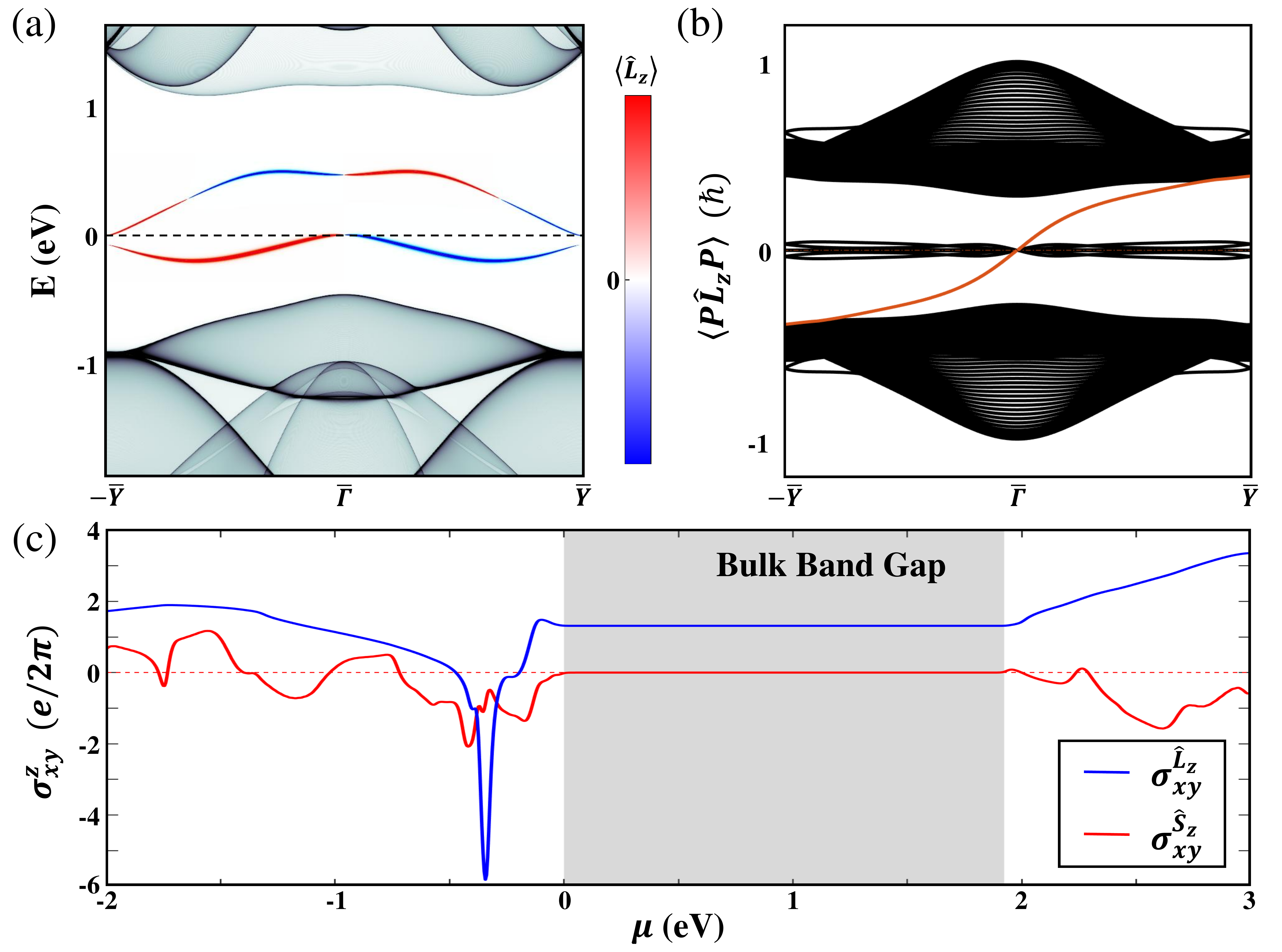}
\caption{(a) Armchair-edge energy band with the orbital texture $\langle\hat{L}_z\rangle$. (b) Edge $\hat{L}_z$-feature spectrum. (c) Orbital (blue line) and spin (red line) Hall conductivity as a function of the chemical potential in monolayer blue-phosphorene.
}
\label{fig:02}
\end{figure}

 According to bulk-boundary correspondence in a Chern insulator, the Chern number corresponds to the number of nontrivial states connecting the valence and conduction bands\cite{hatsugai_chern_1993,mong_edge_2011}. In a feature Chern insulator, these nontrivial edge states can manifest in both the edge energy band and the edge feature spectrum, with the total number of nontrivial states being equal to the feature Chern number\cite{wang_feature_2023,yao_feature-energy_2024}. For monolayer blue-phosphorene, as an OCI, $C_L=1$ thus implies a total of one nontrivial edge state. Figure~\ref{fig:02}(a) presents the armchair edge energy band of monolayer blue-phosphorene, showing floating edge states within the bulk band gap and revealing that the number of nontrivial edge states in the edge energy band is zero, with no states connecting valance and conduction bands. In contrast, the edge $\hat{L}_z$-feature spectrum [Fig.~\ref{fig:02}(b)] displays a nontrivial edge state (orange line) that connect the $\mathbb{I}_+$ and $\mathbb{I}_-$ sectors. The absence of nontrivial states in the edge energy band alongside the presence of a nontrivial state in the edge $\hat{L}_z$-feature spectrum confirms a total of one nontrivial edge state. Notably, since the nontrivial state manifests in the edge $\hat{L}_z$-feature spectrum, the floating edge state seen in Fig.~\ref{fig:02}(a) represents residual feature of the nontrivial state within the edge energy band. The state carries OAM and it is guaranteed by the nonzero orbital Chern number. Unlike the trivial dangling bonds, see SM, the floating edge states cannot be removed by perturbations, distinguishing them as a direct consequence of the orbital Chern topology.

\par For spinful monolayer blue-phosphorene, opposite spin states contribute equally in the orbital degree of freedom  with $C_L=1$ due to the weak SOC, resulting in a orbital Chern number $C_L=2$, see SM for details of calculations with SOC.



The nonzero Chern number suggests the presence of an OHC plateau within the bulk band gap. To investigate the OHE, we employ linear response theory to describe the orbital Hall current as $J_x^{\hat{L}_z}=\sigma_{xy}^{\hat{L}_z}E_y$\cite{bernevig_orbitronics_2005, go_orbitronics_2021, jo_gigantic_2018, go_intrinsic_2018, pezo_orbital_2023, go_orbital_2020, canonico_orbital_2020, bhowal_intrinsic_2020, cysne_disentangling_2021, cysne_orbital_2022, busch_orbital_2023, pezo_orbital_2022, ji_reversal_2024, costa_connecting_2023, yang_monopole-like_2023}, where $\sigma_{xy}^{\hat{L}_z}$ represents the OHC, given by \cite{go_intrinsic_2018, pezo_orbital_2022}

\begin{equation}\label{eq:03}
\begin{split}
    \sigma_{xy}^{\hat{L}_z}=2e\hbar\int_{BZ}\frac{d^2{\bf k}}{(2\pi)^2}\sum_n f_{n}({\bf k})\Omega_{n,xy}^{\hat{L}_z}({\bf k}),
\end{split}
\end{equation}

\noindent where $f_{nk}$ is the Fermi-Dirac distribution and $\Omega_{n,xy}^{\hat{L}_z}({\bf k})$ is the orbital-weighted Berry curvature \cite{go_intrinsic_2018, pezo_orbital_2022}:

\begin{equation}\label{eq:04}
\begin{split}
    \Omega_{n,xy}^{\hat{L}_z}(k)=\sum_{m\neq n}{\rm Im}\frac{\langle u_{nk}|\hat{\mathcal{J}}_x^{\hat{L}_z}|u_{mk}\rangle\langle u_{mk}|\hat{v}_y|u_{nk}\rangle}{(E_{nk}-E_{mk})^2}.
\end{split}
\end{equation}

\noindent Here, $|u_{nk}\rangle$ and $E_{nk}$ denote the Bloch eigenstates and eigenenergies of the Hamiltonian $\hat{\mathcal{H}}({\bf k})$, respectively. The velocity operator along the $x(y)$-direction is, $\hat{v}_{x(y)}=\hbar^{-1} \partial \hat{\mathcal{H}}({\bf k})/\partial k_{x(y)}$, while the OAM current operator is given by $\hat{\mathcal{J}}_x^{\hat{L}_z}=\frac{1}{2}\{\hat{v}_x,\hat{L}_z\}$. Figure~\ref{fig:02}(c) presents the OHC (blue line) for monolayer blue-phosphorene in the presence of SOC. The OHC exhibits a plateau with a height of approximately $1.3 (e/2\pi)$ within the bulk band gap. For a $\hat{L}_z$-conserved Hamiltonian ($[H,\hat{L}_z]=0$) with spherical symmetry, the OHC exhibits a quantized value and it is directly determined by the orbital Chern number, see SM for details. The absence of spherical symmetry, arising from the lattice geometry, leads to a non-$\hat{L}_z$-conserved Hamiltonian and results in a non-quantized OHC plateau. Despite this deviation from quantization, the presence of a nonzero orbital Chern number still guarantees the existence of an OHC plateau. The deviation from perfect quantization in the OHC here is analogous to the non-$\hat{S}_z$-conserved quantum spin Hall effect in SCIs\cite{kane_z_2005}. The spin Hall conductivity (SHC), calculated by replacing the OAM operator with the spin angular momentum operator $\hat{S}_z$\cite{go_intrinsic_2018}, vanishes within the bulk band gap in Fig.~\ref{fig:02}(c), as dictated by the zero spin Chern number, see SM. These OHC and SHC results confirm the presence of an OHC plateau while maintaining a vanishing SHC in monolayer blue-phosphorene.



\begin{figure}[t]
\centering
\includegraphics[width=\linewidth]{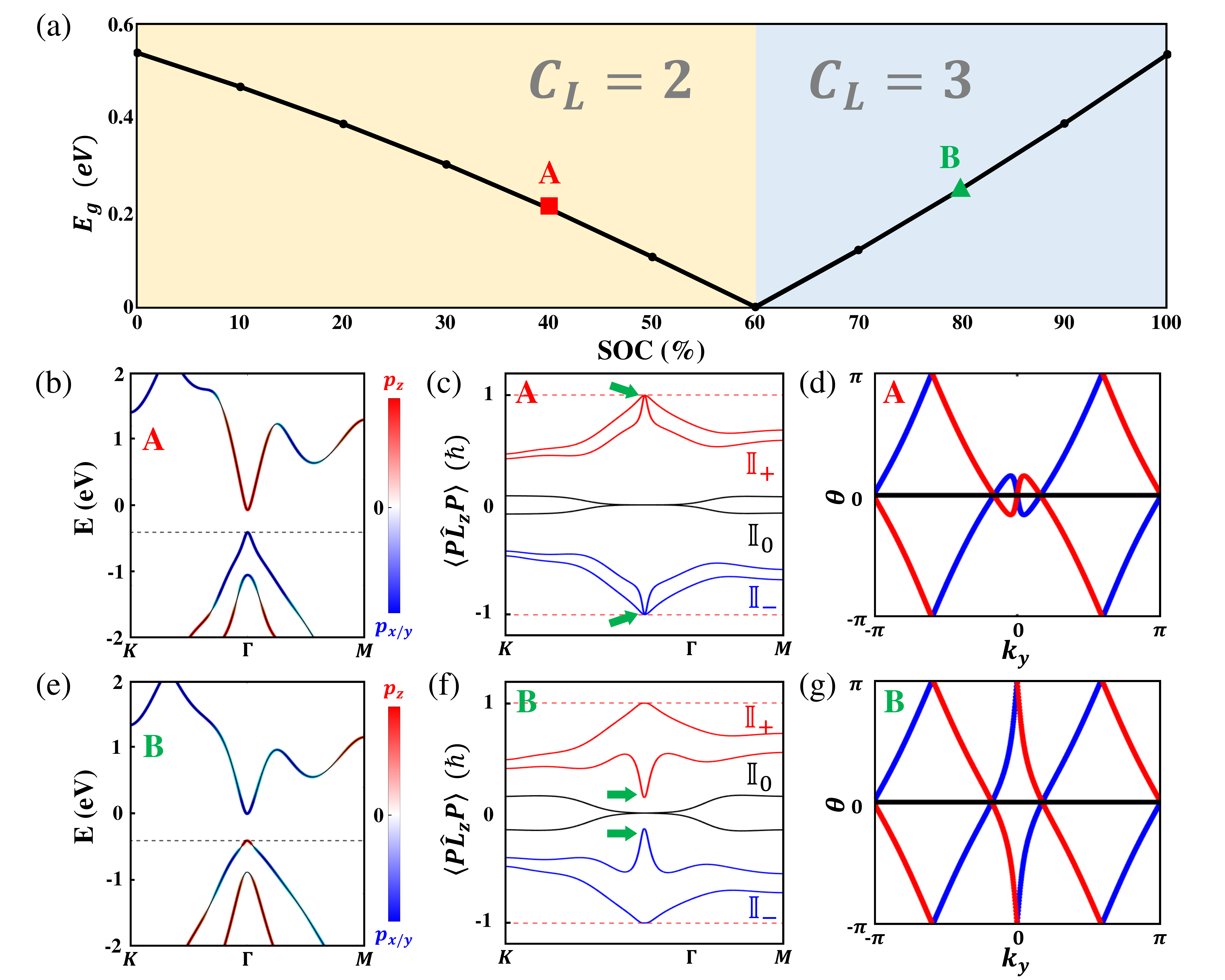}
\caption{(a) Evolution of the band gap at the $\Gamma$ point in monolayer $\beta$-bismuthene as a function of SOC strength. (b,e) Band structure and orbital contributions at points (b) A and (e) B marked in panel (a). (c,f) $\hat{L}_z$-projected feature spectrum at points (c) A and (f) B marked in panel (a). (d,g) $\hat{L}_z$-resolved Wilson loop at points (d) A and (g) B marked in panel (a).
}
\label{fig:03}
\end{figure}

\par Topological phase transitions are of fundamental importance for understanding and manipulating topological properties\cite{bernevig_quantum_2006}. Here, we take SOC as the tuning parameter for driving the topological phase transition in OCIs. Figure~\ref{fig:03}(a) presents the resulting topological phase diagram, illustrating the evolution of the band gap as a function of the SOC strength. These results show that the band gap at the $\Gamma$ point decreases rapidly with increasing SOC strength, and undergoes a phase transition around 60$\%$ of the SOC strength in the pristine compound. Before this phase transition, at point A, the orbital contributions exhibit a band inversion, see Fig.~\ref{fig:03}(b) where the $\hat{L}_z$-projected feature spectrum [Fig.~\ref{fig:03}(c)] and the $\hat{L}_z$-projected Wilson's loop [Fig.~\ref{fig:03}(d)] indicate a total orbital Chern number of $C_L=2$, which closely resembles the topological properties of monolayer blue-phosphorene. After the phase transition at point-B, an additional band inversion occurs in which the $p_z$ orbital inverts from the valence band [Fig.\ref{fig:03}(b)] to conduction band [Fig.\ref{fig:03}(e)] around the $\Gamma$ point. Two valence bands, previously dominated by $p_{x/y}$ orbitals, are replaced by the $p_z$ orbital, inducing a modification of $\hat{L}_z$-feature spectrum. As a result, the feature bands in $\mathbb{I}_{\pm1}$ sectors [green arrows in Fig.~\ref{fig:03}(c)] shift closer to 0 [green arrows in Fig.~\ref{fig:03}(f)]. This transition modifies the ground state in the orbital degree of freedom, leading to a change in the orbital Chern number, which increases from $C_L=2$ [Fig.~\ref{fig:03}(d)] to $C_L=3$ [Fig.~\ref{fig:03}(g)]. Specifically, the SOC-driven topological phase transition in OCIs could be achieved by adding heavier elements in honeycomb pnictogen monolayers, such as bismuth. The phase transition is accompanied by a change in spin Chern number, see SM, where the resulting SCI phase aligns with previous studies confirming that monolayer $\beta$-bismuthene is a two-dimensional SCI in the presence of SOC.\cite{zhou_epitaxial_2014, wada_localized_2011, drozdov_one-dimensional_2014, akturk_single_2016, liu_stable_2011}.


\begin{figure}[t]
\centering
\includegraphics[width=\linewidth]{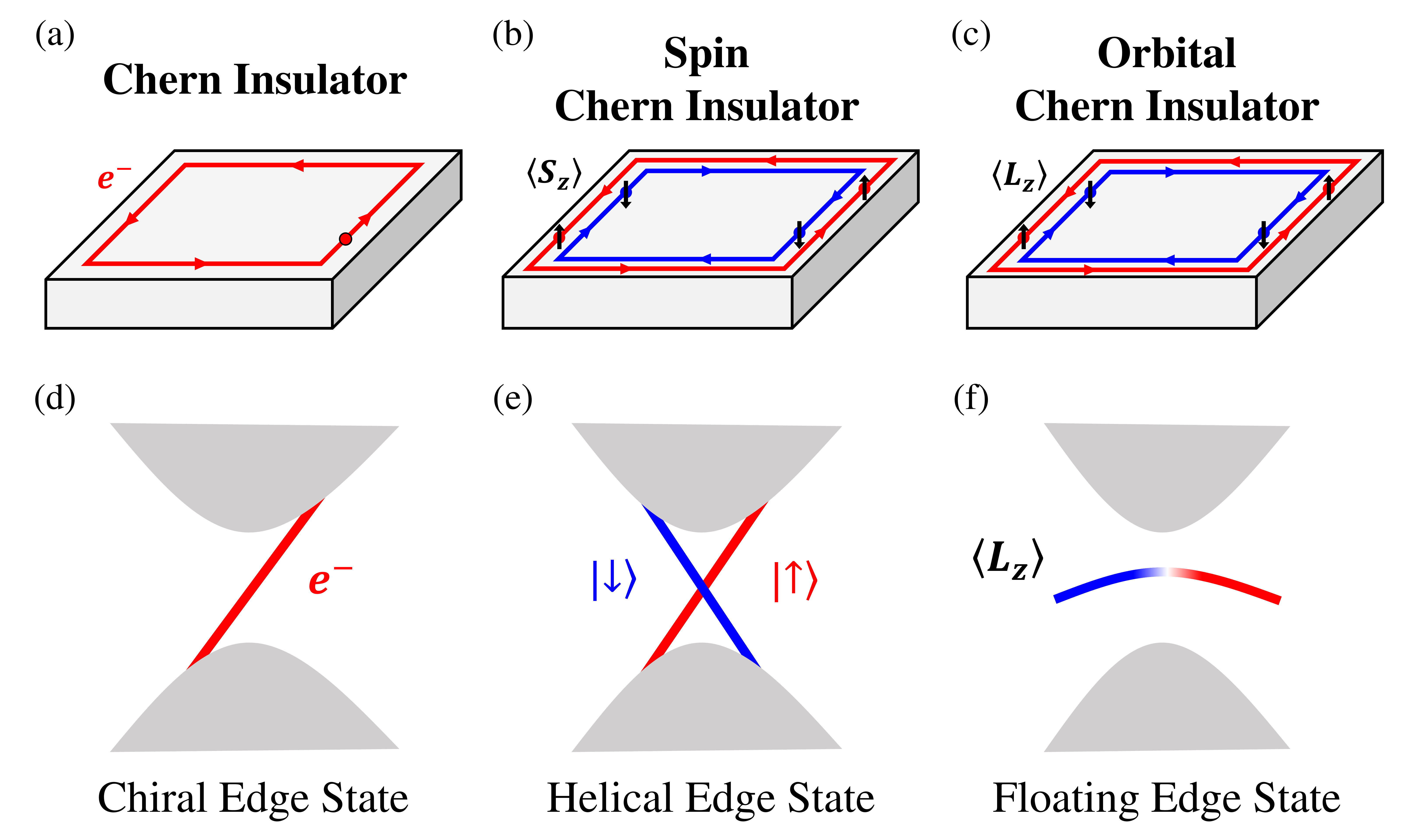}
\caption{(a-c)  Edge transport behavior for (a) CIs, (b) SCIs, and (c) OCIs. (d-f) Edge energy band structures corresponding to (d) chiral edge states in CIs, (e) helical edge states in SCIs, and (f) floating edge states in OCIs.
}
\label{fig:04}
\end{figure}

\par \textit{Discussion---} We summarize the topological phases of quantum matter characterized by Chern-like invariants across the charge, spin, and orbital degrees of freedom in Fig.~\ref{fig:04}. The CI represents the charge degree of freedom, where the edge signatures in Fig.~\ref{fig:04}(a) demonstrate charge transport along the edge, corresponding to the chiral edge states shown in Fig.~\ref{fig:04}(d). The Chern number, as the defining topological invariant, can be experimentally measured through the quantized anomalous Hall conductivity\cite{chang_experimental_2013,deng_quantum_2020}. In the spin degree of freedom, SCIs exhibit spin transport along the edge [Fig.~\ref{fig:04}(b)]. In this phase, spin-up and spin-down currents flow in opposite directions, leading to the formation of helical edge states [Fig.~\ref{fig:04}(e)]. A nonzero spin Chern number corresponds to a SHC plateau within the bulk band gap. In the orbital degree of freedom, OCIs exhibit edge transport behavior analogous to SCIs, with the key distinction that charge carriers flow in opposite directions while carrying opposite OAM [Fig.~\ref{fig:04}(c)], rather than spin. Notably, OCIs such as monolayer blue-phosphorene exhibit floating edge states [Fig.~\ref{fig:04}(f)] that emerge as residual features of nontrivial states in the edge $\hat{L}_z$-feature spectrum. This results in a unique bulk-boundary correspondence, distinguishing OCIs from conventional topological phases. Like the SCIs, the nonzero orbital Chern number also suggests the presence of an OHC plateau within the bulk band gap.

\par The presence of floating type edge states should be noted in obstructed atomic insulators\cite{po_fragile_2018,khalaf_boundary-obstructed_2021,xu_three-dimensional_2021,xu_filling-enforced_2021,li_floating_2025,verma_atomically_2024,liu_observation_2024,li_obstructed_2022,liu_spectroscopic_2023}, including puckered black-phosphorene\cite{verma_atomically_2024}, and theoretically proposed altermagnets\cite{li_floating_2025}. In obstructed atomic insulators, the charge from the floating edge states is localized at specific Wyckoff position, protected by a symmetry-based real space invariant. Our approach, based on the feature spectrum topology, provides a framework for demonstrating the topological nature of these experimentally observed floating edge states\cite{ liu_observation_2024,li_obstructed_2022,liu_spectroscopic_2023}, extending beyond symmetry-based classification. Furthermore, our findings suggest the possibility of measurable orbital Hall responses in these materials, providing new insights into their transport properties. 


\par \textit{Conclusion---} Using the feature spectrum topology approach, we have presented a systematic framework for investigating OCIs, which are a new topological phase of quantum matter driven by the orbital degree of freedom. Through a systematic analysis of monolayer blue-phosphorene as a representative example, we provide new insight into the nature of orbital-driven nontrivial topology in materials previously classified as normal (trivial) insulators\cite{ tani_topological_2022,ezawa_topological_2014,rudenko_electronic_2017,zhang_topological_2012,chuang_tunable_2013,kamal_arsenene_2015}. Furthermore, by using the strength of the SOC as a tuning parameter, we discuss how the OCI phases could be manipulated, highlighting the potential for engineering candidate materials. Our study establishes OCIs as a distinct topological phase of quantum matter and paves the way for exploration of orbital-driven quantum phenomena and driving future advancements in orbitronics.




\par \textit{Acknowledgment---} We sincerely thank Rahul Verma, Shin-Ming Huang, and Yann-Wen Lan for very valuable discussions. T.-R.C. was supported by National Science and Technology Council (NSTC) in Taiwan (Program No. MOST111-2628-M-006-003-MY3 and NSTC113-2124-M-006-009-MY3), National Cheng Kung University. (NCKU), Taiwan, and National Center for Theoretical Sciences, Taiwan. This research was supported, in part, by the Higher Education Sprout Project, Ministry of Education to the Headquarters of University Advancement at NCKU. T.-R.C. thanks the National Center for High-performance Computing (NCHC) of National Applied Research Laboratories (NARLabs) in Taiwan for providing computational and storage resources. H.L. acknowledges the support by Academia Sinica in Taiwan under grant number AS-iMATE-113-15. The work at Northeastern University was supported by the National Science Foundation through the Expand-QISE award NSF-OMA-2329067 and benefited from the resources of Northeastern University’s Advanced Scientific Computation Center, the Discovery Cluster, the Massachusetts Technology Collaborative award MTC-22032, and the Quantum Materials and Sensing Institute.


%

\end{document}